\documentclass{article}
\usepackage{spconf,amsmath,graphicx}
\usepackage{algorithm}
\usepackage{algorithmic}
\usepackage{amsfonts,amssymb}
\usepackage{hhline}
\usepackage{footmisc}

\title{Hierarchical Softmax for End-to-End Low-resource Multilingual Speech Recognition}
\name{\em{Qianying Liu${}^{1\ast}$, Zhuo Gong${}^{3\ast}$, Zhengdong Yang${}^{1\ast}$,  Yuhang Yang${}^{2}$, Sheng Li${}^{4}$, Chenchen Ding${}^{4}$}, \\\em{
Nobuaki Minematsu${}^{3}$, Hao Huang${}^{2}$, Fei Cheng${}^{1}$, Chenhui Chu${}^{1}$, Sadao Kurohashi${}^{1}$}\thanks{$\ast$ denotes equal contribution.}}
\address{${}^{1}$Graduate School of Informatics, Kyoto University, Sakyo-ku, Kyoto, Japan \\
${}^{2}$School of Information Science and Engineering, Xinjiang University, Urumqi, China \\
${}^{3}$Department of EEIS, Graduate School of Engineering, The University of Tokyo, Tokyo, Japan \\
${}^{4}$National Institute of Information and Communications Technology (NICT), Kyoto, Japan
}

\begin{document}
%

\maketitle
\begin{abstract}
Low-resource speech recognition has been long-suffering from insufficient training data. 
In this paper, we propose an approach that leverages neighboring languages to improve low-resource scenario performance, founded on the hypothesis that similar linguistic units in neighboring languages exhibit comparable term frequency distributions, which enables us to construct a Huffman tree for performing multilingual hierarchical Softmax decoding. This hierarchical structure enables cross-lingual knowledge sharing among similar tokens, thereby enhancing low-resource training outcomes. Empirical analyses demonstrate that our method is effective in improving the accuracy and efficiency of low-resource speech recognition.
\end{abstract}

\section{Introduction}
\label{sec:intro}

Automatic speech recognition (ASR) systems have gained remarkable progress in the past few years. Nevertheless, the present ASR systems cater to only approximately 100 out of the 7000 spoken languages worldwide. To address this limitation, multilingual models have garnered much attention. These models can learn universal features, transferable from resource-rich to resource-limited languages, and support multiple languages with a single ASR model.
Early studies utilized context-dependent deep neural network hidden Markov models \cite{dnnhmm}, which relied on hand-crafted pronunciation lexicons. However, when adapted to low-resource languages, such systems exhibit limitations due to the absence of sufficient modeling techniques. Attention-based end-to-end (E2E) models simplify training and eliminate the dependence on pronunciation lexicons~\cite{hinton2012deep,graves2014towards,bahdanau2016end}.
Recent studies employing E2E models have focused on learning universal representations at the encoding stage, using transfer learning techniques~\cite{m12, m13, li2019end, m15, m16, m17}, as well as on hierarchical embedding of phonemes, phones, and phonological articulatory attributes~\cite{li2021hierarchical}. Meanwhile, large pretrained models \cite{m3, m4, m5, m6} and multilingual speech corpora \cite{m7, m8, m9, cmu-data, m11} have been investigated for learning pretrained representations.

In this paper, we aim to investigate the explicit transfer of cross-language knowledge at the decoding stage, which has been largely unexplored in prior studies that focus on encoder representations. Based on linguistic studies on the presence of similar modeling unit distributions in neighboring languages~\cite{artetxe-etal-2018-robust}, we propose an efficient method to capture the similarity among these units, such as characters and globalphones, at the decoding stage.
Our approach utilizes Huffman coding to automatically capture the similarity among modeling units, relying only on monolingual data. We introduce hierarchical Softmax (H-Softmax) \cite{mikolov2013distributed}, an approximation softmax inspired by binary trees, to model the similarity during decoding. This structure enables similar units to share decoder representations, thus improving model performance, and also breaks down the expensive softmax step into several binary classifications, thereby enhancing model efficiency~\cite{mohammed2018effectiveness}. We design a vectorization algorithm that can expedite the training and inference procedures, enabling our method to outperform the vanilla softmax on GPUs in terms of efficiency.

With the combination of the two components, our method:
\begin{itemize}
    \item Automatically captures cross-lingual modeling unit similarity for multilingual ASR.
    \item  Leverages H-Softmax to achieve higher efficiency and reduce computational complexity.
\end{itemize}

While previous studies has utilized H-Softmax in the neural language model of ASR~\cite{m1, m2}. However, to the best of our knowledge, no existing work has investigated the direct application of H-Softmax in the acoustic model. Furthermore, our study is the first to explore the potential of H-Softmax for low-resource multilingual ASR.



\begin{figure}[htb]
  \setlength{\abovecaptionskip}{15pt}
  \centering
  \includegraphics[width=0.5\textwidth]{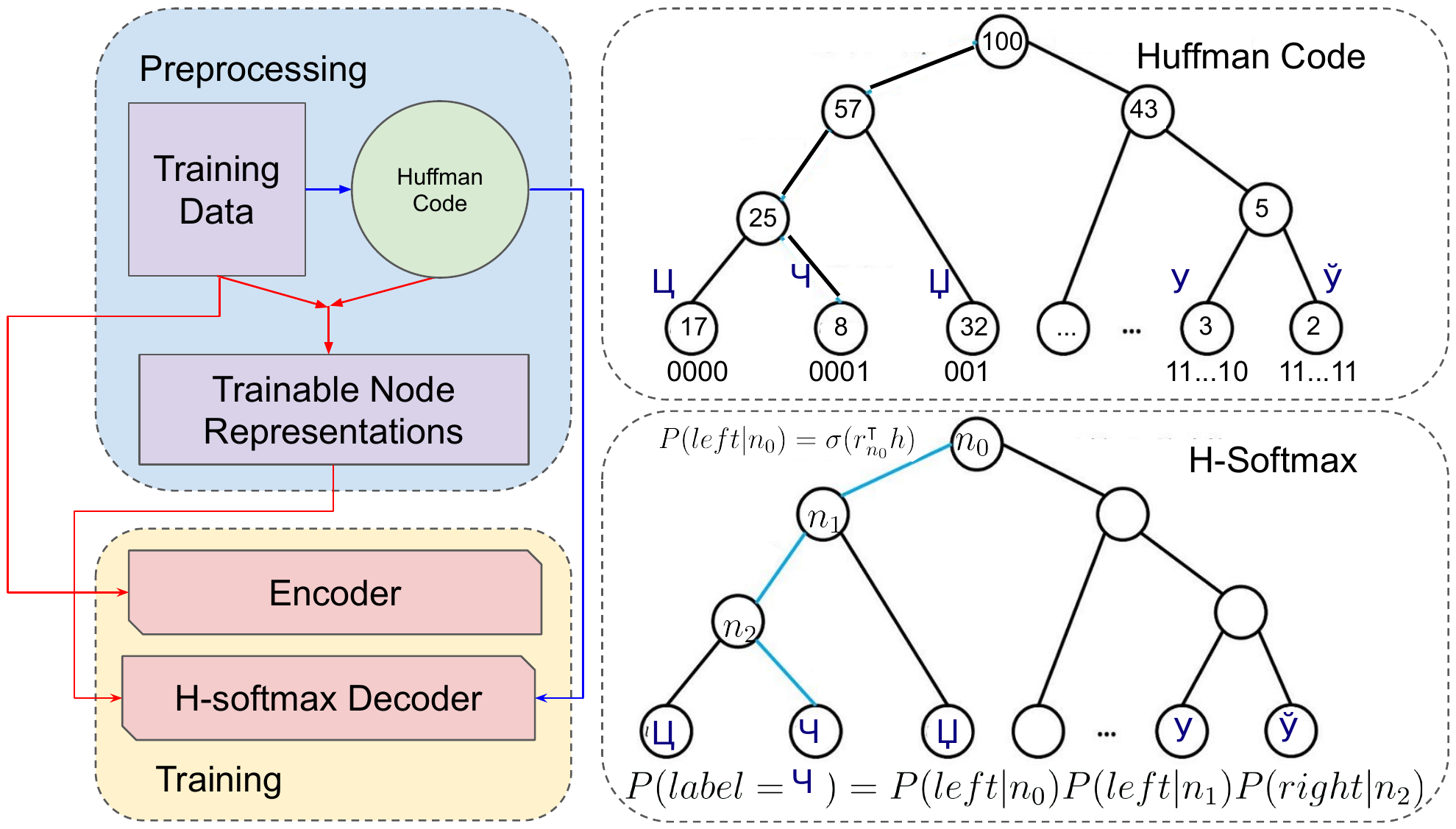}
  \caption{The flowchart of the proposed method. The blue lines stands for determination relation, and the red line stands for how the data goes through the model. }
  \label{fig:proposed}
\end{figure}

\section{Proposed Method}
\label{sec:method}

In this section, we introduce the proposed method of this paper, as shown in Fig \ref{fig:proposed}. We first use the training text data to determine the Huffman code for each token as described in subsection \ref{ssec:hc}. Then we build the ASR model, where an encoder takes in the source speech signals, and the decoder predicts the Huffman code of target text with H-Softmax. At the inference stage, the model predicts a sequence of post-processed Huffman codes to text.


\subsection{Huffman Code}
\label{ssec:hc}

Based on the assumption that neighbour languages share similar token distribution, the concept of our proposed method is to generate a representation code for each token via frequency clustering. We first generate the Huffman code of each token.
Formally, given the multilingual token vocabulary $V=\{t_1, t_2, ...t_N\}$, where $N$ is the vocabulary size, we maintain the term frequency set $S_p=\{p_{t_i}\}_{i=1}^N$, where the same token in different languages are considered as one token. 
With $S_p$, we generate a Huffman tree of $V=\{t_i\}$ via frequency clustering and further recursively generate the Huffman code
by assigning 0 to the left subtree and 1 to the right subtree. 

\subsection{Model Architecture} 

For the ASR model, we use conformer~\cite{gulati2020conformer} as the encoder and transformer~\cite{vaswani2017attention} as the decoder. For the decoder, we replace the vanilla softmax with H-Softmax.

H-Softmax organizes the output vocabulary into a tree where the leaves are the vocabulary tokens, and the intermediate nodes are latent variables \cite{mikolov2013distributed}. We use the Huffman tree generated in subsection \ref{ssec:hc} as the tree for H-Softmax that there is a unique path from the root to each token, forming a complete binary tree. 
We follow \cite{mikolov2013distributed} and apply the binary tree H-Softmax. The decoding procedure is transformed into predicting one leaf node of the binary tree at each timestep. Each leaf node, which represents a token, could be reached by a path from the root through the inner nodes. Given the transformer output hidden state $h$ and trainable node representation vectors $\{r_i\}$, the final possibility of a leaf node $w_i$ could be represented as:

\vspace{-12pt}
\begin{equation}
\begin{aligned}
        P(label = {w}_i) = \prod^{{path}} P({path}_k|n_k) \\
        = \prod^{{path}}
        \begin{cases}
        \sigma (r_k^\intercal h) & \text{if  } {path}_k = left, \\
        1-\sigma (r_k^\intercal h) & \text{if  }  {path}_k = right. 
        \end{cases}
\end{aligned}
\label{eq:prob}
\end{equation}

\noindent where $n_k$ stands for the $k$th node on the path from the root to ${w}_i$ such as $n_0$, $n_1$ and $n_2$ in Fig.~\ref{fig:proposed}; ${path}_k$ stands for the branch leading towards ${w}_i$ which is $left$ or $right$, and $\sigma(\cdot)$ stands for sigmoid function.

\subsection{Efficient Implementation of H-Softmax}
\label{ssec:HS}
While decomposing the Softmax to a binary tree H-Softmax reduces the decoding time complexity from O(V) to O(log(V)), and the train time complexity remains O(Vlog(V)). Since previous H-Softmax implementation\cite{mohammed2018effectiveness} is on CPU, considering the order of magnitude difference between CPU's and GPU's FLOPs, the challenge of improving the efficiency of model training lies in designing an implementation of H-Softmax for GPU training. We propose a vectorization algorithm to accelerate training and decoding procedures on GPU. 

\begin{figure}[H]
    \vspace{-10pt}   
     \includegraphics[width=1\columnwidth]{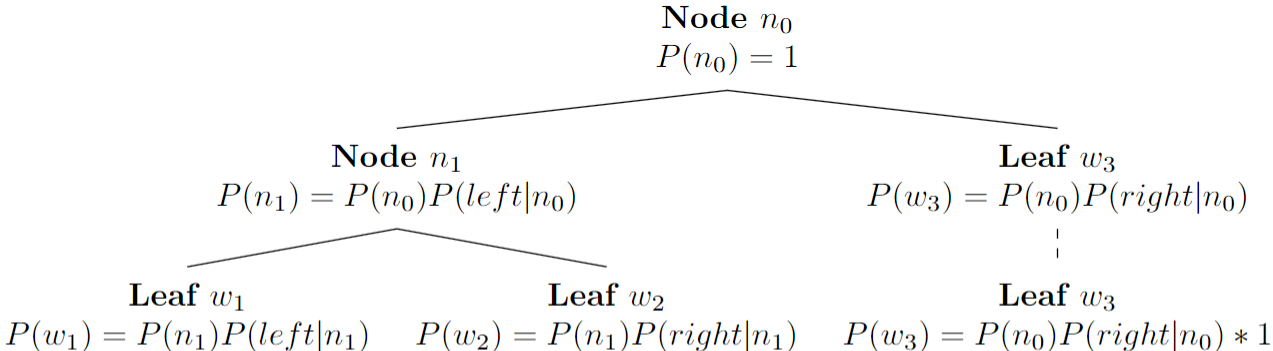}
    \vspace{-20pt}     
    \caption{A typical H-Softmax tree structure. Leaf $w_3$ has a virtual child with the same probability of aligning each leaf node to the same depth, so it is conceptually possible for path vectorization.}
    \label{fig:tree}
    \vspace{-5pt}   
\end{figure}
To explain the core idea of our vectorization algorithm, we show a typical H-Softmax tree structure in Fig~\ref{fig:tree}. Then, we can vectorize log-probability calculations from Eq~\ref{eq:prob} to the followings:
\begin{align*}
&\begin{bmatrix}
\log P(w_1)\\ 
\log P(w_2)\\ 
\log P(w_3)
\end{bmatrix}
= \begin{bmatrix}
\log[\sigma(r_1^\intercal h)\sigma(r_2^\intercal h)]\\ 
\log[\sigma(r_1^\intercal h)(1-\sigma(r_2^\intercal h))]\\ 
\log[1-\sigma(r_1^\intercal h)]
\end{bmatrix} \\
&=\sum\limits_{column}\log\begin{bmatrix}
1*\sigma(r_1^\intercal h)+0 & 1*\sigma(r_2^\intercal h)+0\\ 
1*\sigma(r_1^\intercal h)+0 & -1*\sigma(r_2^\intercal h)+1\\ 
-1*\sigma(r_1^\intercal h)+1 & 0*\sigma(r_2^\intercal h)+1
\end{bmatrix} \\
&=\sum\limits_{column} \log(Sign \circ \sigma(p) + Bias)
\label{eq:log_prob}
\end{align*}
where $Sign$ is a 3 by 2 matrix of $\sigma$s' signs, $Bias$ is a 3 by 2 matrix of $\sigma$s' biases and
$p$ is the result vector of the inner product between node vectors and $h$.
After building the Huffman tree, the $Sign$ and $Bias$ matrices are fixed. So in the training stage, leaf node log probabilities can be acquired only by vector operations. 

For decoding, we only need leaves with the highest probabilities. To directly calculate this objective, different from training, we also develop a path-encoding-based multi-layer beam searching on GPU for H-softmax\footref{github} to retain the time efficiency advantage of time-space complexity O(log(V)) compared to vanilla softmax's O(V).

\section{Experiment Evaluations}
\label{sec:review}

In this section, we evaluate the proposed method in two low-resource settings. To examine the effectiveness of our method in more extensive settings, specifically in the same language group and cross-language groups, we first simulate a low resource setting by down-sampling from a speech-to-text multilingual large-scale dataset. To further verify the performance on natural low-resource languages and other token modeling units, we test our method on an extremely low-resourced zero-shot cross-lingual speech-to-phoneme setting. For the high-resource setting, where every token could be fully trained, modeling of neighbors could no longer be useful, and thus our experiments focus on the low-resource setting.


\subsection{Data Description}
\label{ssec:chineseasr}

We sample our speech-to-text datasets from Common Voice Corpus 11.0.
We selected three different linguistic groups: Romance, Slavic, and Turkic. For each group, we selected five languages from the corpus and constructed training and testing sets with the validated data of each language.
With the data size of many existing datasets being around $20{\sim}30$ 
hours~\cite{DBLP:journals/corr/WangZ15e, THUYG-20}, we downsampled the training data to the extent of 30 hours per language on average to simulate a low-resource setting. As a result, the total size of training data changed from 5492 hours to 450 hours, with the general downsampling ratio $\lambda=0.082$. 
Different downsampling ratios are used for different languages to counter the imbalance of data size among languages.
For a set of languages $\{L_1, ...,L_m\}$ with their proportion $\{p_1, ...,p_m\}$ , the downsampling ratio for language $L_i$ is obtained by $ \lambda_i = \frac{p_i^{\alpha-1}}{\sum_{j=1}^{m} p_j^\alpha}\lambda$,  where $\alpha$ is a smooth coefficient that we set to 0.5.
The size of testing sets is the lesser of 10\% of the validated data and 10,000 utterances.

For speech-to-phoneme datasets, we use UCLA Phonetic Corpus \cite{cmu-data}. The Edo-Ancient Tokyo (bin), Kele-Congo (sbc), Malayalam-Malay (mal),  Creole-Cape Verde (kea), Klao-Liberia (klu), Arabic-Tunisian Spoken (aeb), Makasar-South Sulawesi (mak), Finnish-Finland (fin), Abkhaz-North Caucasian (abk) and Aceh-Sumatra (ace) are testing languages and other languages are used for training. 



\begin{table}[ht] 
\setlength{\abovecaptionskip}{10pt}
\begin{center}
{
\caption{Speech-to-text datasets statistics.}
\label{tab:textd}
\begin{tabular}{|l|l||r|r|}
\hline
Group & Language & Training & Testing  \\
 &  & (Hours) & (Hours)  \\
\hhline{|=|=|=|=|}
 & Catalan (ca) & 76.3  & 15.3  \\
\cline{2-4}
 & Spanish (es) & 37.6  &  14.4  \\
\cline{2-4}
Romance & French (fr) & 55.7  &  13.4  \\
\cline{2-4}
 & Italian (it) & 33.4  &  15.0  \\
\cline{2-4}
 & Portugal (pt) & 20.7  &  11.4  \\
\hline
 & Belarusian (be) & 63.0  &  13.9  \\
\cline{2-4}
 & Czech (cs) & 42.5  &  5.8  \\
\cline{2-4}
Slavic & Polish (pl) & 37.8  &  12.3  \\
\cline{2-4}
 & Russian (ru) & 27.3  &  14.5  \\
\cline{2-4}
 & Ukrainian (uk) & 23.4  &  7.2  \\
\hline
 & Bashkir (ba) & 29.6  &  12.6  \\
\cline{2-4}
 & Kyrgyz (ky) & 11.3  &  3.8  \\
\cline{2-4}
Turkic & Turkish (tr) & 16.9  &  8.4  \\
\cline{2-4}
 & Tatar (tt) & 10.0  &  3.0  \\
\cline{2-4}
 & Uzbek (uz) & 18.1  &  10.4  \\
\hline

\hline
\end{tabular}
}\\
\end{center}
\vspace{-20pt}
\end{table}

\begin{table}[ht] 
\setlength{\abovecaptionskip}{10pt}
\begin{center}
{
\caption{Speech-to-phoneme datasets.}
\label{tab:phoned}
\begin{tabular}{|l|l||c|}
\hline
Language & Dataset & Hours  \\
\hhline{|=|=|=|}
UCLA Phonetic Corpus  & Training (87 lang.) &  1.86    \\
\cline{2-3}
\mbox{(97 languages)} & Testing (10 lang.)  &  0.14  \\
\hline
\end{tabular}
}\\
\end{center}
\vspace{-20pt}
\end{table}

\begin{table*}[ht] 
\setlength{\abovecaptionskip}{10pt}
\begin{center}
{
\caption{PER\% on speech-to-phoneme datasets.}
\label{tab:phone}
\begin{tabular}{|l||r|r|r|r|r|r|r|r|r|r|r|}
\hline
Model & bin & sbc & mal & kea & klu & aeb & mak & fin & abk & ace & Overall  \\
\hhline{|=|=|=|=|=|=|=|=|=|=|=|=|}
Softmax  &  70.4 &  87.4  &   98.0  &   83.4  &   86.2  &   84.1&   84.5  &   94.2  &  75.0  &   75.5 & 85.2 \\
\hline
H-Softmax  &  \textbf{38.0}  &  \textbf{68.9}   &  \textbf{80.8}   &  \textbf{59.6}   &  \textbf{62.9}   &  \textbf{60.9}&  \textbf{73.7}   &   \textbf{75.2}   &  \textbf{70.9}   &  \textbf{48.6}   &  \textbf{64.7} \\
\hline
\end{tabular}
}\\
\end{center}

\vspace{-20pt}
\end{table*}

\begin{table}[ht] 
\setlength{\abovecaptionskip}{10pt}
\begin{center}
{
\caption{CER\% on speech-to-text datasets. Models are trained with languages within the same language group.}
\label{tab:text-f}
\begin{tabular}{|l||r|r|r|r|r|r|}
\hline
Model & ca & es & fr & it & pt & Overall   \\
\hhline{|=|=|=|=|=|=|=|}
Softmax  &  5.5&	9.4&	12.1&	8.8&	\textbf{9.7} & 9.1\\
\hline
H-Softmax  &  \textbf{5.3}&	\textbf{8.8}&	\textbf{11.3}&	\textbf{8.4}	& 10.0   & \textbf{8.8}  \\
\hline
\hline
Model & be & cs & pl & ru & uk & Overall   \\
\hhline{|=|=|=|=|=|=|=|}
Softmax  &  5.0&	5.4&	9.2&	11.6&	\textbf{11.5}  &  8.5 \\
\hline
H-Softmax  &  \textbf{4.8}&	5.4&	\textbf{8.8}&	\textbf{10.4}&	11.6   &  \textbf{8.2} \\
\hline
\hline
Model & ba & ky & tr & tt & uz & Overall   \\
\hhline{|=|=|=|=|=|=|=|}
Softmax  &  16.9&18.3	&	14.0&	10.2&	20.6& 16.0 \\
\hline
H-Softmax  & \textbf{14.4}&	\textbf{17.2}&	\textbf{12.8}&	\textbf{9.1}&	\textbf{16.3}& \textbf{14.0}	\\
\hline
\end{tabular}
}\\
\end{center}
\vspace{-20pt}
\end{table}

\subsection{Model Training}
For acoustic features, the 80-dimensional log-Mel filterbanks (FBANK) are computed 
with a 25ms window and a 10ms shift. Besides, SpecAugment \cite{specaug} is applied to 2 frequency masks with maximum frequency mask (F = 10) and 2 time masks with maximum time mask (T = 50) to alleviate over-fitting. 

Both H-Softmax and Softmax models are trained using the same network structure. The networks are constructed using WeNet toolkit \cite{yao21_interspeech}.\footnote{Our implementation is  https://github.com/Derek-Gong/hsoftmax\label{github}} Two convolution sub-sampling layers with kernel size 3$\times$3 and stride 2 are used in the front of the encoder. For model parameters, we use 12 conformer layers for the encoder and 6 transformer layers for the decoder. 

Adam optimizer is used with a learning rate schedule with 25,000 warm-up steps. The initial learning rate is 0.00005. 100 max epochs for speech-to-text datasets and 80 max epochs for speech-to-phoneme datasets.  


\subsection{Speech-to-text Recognition Evaluation}

We conducted two sets of experiments on speech-to-text datasets: training with languages within one language group and training with languages across language groups. 
We tokenized the transcriptions at the character level as we verified that it outperforms tokenizing in the sub-word level (such as BPE).\footnote{Preliminary experiments on Slavic language group with traditional Softmax results in a CER of 8.5\% for character-level tokenization, 8.9\% for BPE (vacab size = 500) and 9.6\% for BPE (vacab size = 5000).}
As shown in Table \ref{tab:text-f}, when training with languages within the same language group, our Huffman code achieves better performance (character error rate, CER\%) than traditional Softmax for all three groups, which demonstrates the effectiveness of our method.
Results in Table \ref{tab:text-all} show that our method can also make improvements when training with the combined data in 15 languages across different language groups, showing that our method works in a more extensive range of scenarios than we expected.
In addition, our method is more robust as there are no languages with a distinctively high error rate, unlike traditional Softmax.


\subsection{Zero-shot Cross-lingual Speech-to-phoneme Recognition Evaluation}

Our proposed model demonstrated superior performance compared to the conventional model (phone error rate, PER\%) across all languages on the UCLA Phonetic corpus, as presented in Table \ref{tab:phone}. Overall, our approach outperformed the softmax baseline by a significant margin of 20.51\% PER. On the `bin' language, the performance gap between softmax and H-softmax was 32.33\%. These results showcase the effectiveness of our method in automatically constructing universal representations for multilingual ASR and achieving zero-shot cross-lingual phoneme recognition.


\subsection{Decoding Speed}


We observed decoding acceleration with our proposed H-Softmax model in Table \ref{tab:speed}. Decomposing the Softmax output layer to a binary tree reduces the complexity of obtaining probability distribution from O(V) to O(log(V)), which leads to improvement in efficiency. The results also show that the sentence length (tokens per sentences) determines the decoding time difference between Softmax and H-softmax. Longer sentences take more time steps for inference, and the time difference between the two models exponentially increases.  \begin{table}[t] 
\setlength{\abovecaptionskip}{10pt}
\begin{center}
{
\caption{CER\% on speech-to-text datasets. Models are trained with all the languages across language groups.}
\label{tab:text-all}
\begin{tabular}{|l||r|r|r|r|r|r|}
\hline
Model & ca & es & fr & it & pt & Overall   \\
\hhline{|=|=|=|=|=|=|=|}
Softmax  &  7.3&	9.4&	15.1&	8.4&	21.8 & 12.4\\
\hline
H-Softmax  &  \textbf{5.7}&	\textbf{9.0}&	\textbf{11.3}&	\textbf{7.3}	& \textbf{15.4}   &  \textbf{9.7} \\
\hline
\hline
Model & be & cs & pl & ru & uk & Overall   \\
\hhline{|=|=|=|=|=|=|=|}
Softmax  &  \textbf{5.8}&	\textbf{5.8}	&8.5&	10.6&	11.5  &  8.4  \\
\hline
H-Softmax  &  6.1&	5.9&	\textbf{8.2}&	\textbf{8.5}&	\textbf{9.8}   & \textbf{7.7}  \\
\hline
\hline
Model & ba & ky & tr & tt & uz & Overall   \\
\hhline{|=|=|=|=|=|=|=|}
Softmax  &  11.8&	13.7&	12.3&	7.6&	\textbf{13.5}& 11.8	 \\
\hline
H-Softmax  & \textbf{11.0}&	\textbf{13.6}&	\textbf{10.3}&	\textbf{7.1}&	13.7&	\textbf{11.1} \\
\hline
\end{tabular}
}\\
\end{center}

\vspace{-20pt}
\end{table}

\begin{table}[t] 
\setlength{\abovecaptionskip}{10pt}
\begin{center}
{
\caption{RTF (real-time factor) of the decoding process of Table \ref{tab:text-all}. 3 languages are selected to show the decoding speed with different tokens 
per sentences (tok/sent). }
\label{tab:speed}
\begin{tabular}{|l||r|r|r|r|r|}
\hline
Model & tr & pl & ru    \\
 & 32.6 tok/sent & 47.4 tok/sent & 63.1 tok/sent    \\
\hhline{|=|=|=|=|}
Softmax  &  0.022 &  0.023  &   0.026      \\
\hline
H-Softmax  &  \textbf{0.018}  &  \textbf{0.018}   &  \textbf{0.020}    \\
\hline
\end{tabular}
}\\
\end{center}
\vspace{-20pt}
\end{table}

\section{Conclusion}

This paper proposes an automatic method to generate cross-lingual representations, which facilitate the acquisition of cross-lingual universal features in neighboring languages. Our approach employs Huffman coding, which utilizes token frequencies (characters, globalphones, etc.) to bridge different languages. Furthermore, we introduce H-Softmax, which improves model performance by enabling similar units that share Huffman binary representations and accelerates decoding compared to traditional Softmax methods.
As future work, we aim to design binary codes that incorporate not only shallow frequency terms but also more semantically meaningful features, such as token embeddings.

\section{Acknowledgements}
This study was supported by JST SPRING Grant No.JPMJSP2110 and Grant-in-Aid for Scientific Research (C) No. 23K11227.



\bibliographystyle{IEEEbib}\small
\bibliography{mybib, refs}

\clearpage

\appendix

\section{Huffman Coding} 
\label{ssec:Huffman}

Formally, given a set of $m$ languages $\{L_i\}_{i=1}^m$ and the corresponding character sets $S^i=\{c^i_1, c^i_2, ...c^i_{V^i}\}$, where $V^i$ is the character vocabulary size. For each language $L_i$  individually, we maintain the term frequency $S^i_p = \{p_{c^i_j}\}$ of each character $j$ in monolingual data. Term frequency is defined as follow:

\begin{equation}
    p_{c^i_j} = \frac{f_{c^i_j}}{\sum_{j=1}^{V^i}{f_{c^i_j}}}
\end{equation}

Where $f_{c^i_j}$ is the raw count of a term $c^i_j$ in the monolingual data.

\begin{algorithm}
\begin{algorithmic}
\REQUIRE $S_p$
\STATE Assign leaf nodes $\{N_{c^i_j}\}$ for elements in $S_p$;
\STATE sort $\{N_{c^i_j}\}$ based on $\{p_{c^i_j}\}$ order to form a priority queue $Q$;
\WHILE{$sizeof(Q)\geq 2$}
\STATE Remove 2 lowest probability nodes $N_a$ and $N_b$ from $Q$;
\STATE Create a new internal node $N_c$ with $N_a$ and $N_b$ as children, probability $p_c \gets$ sum of $p_a$ and $p_b$;
\STATE Add $N_c$ to $Q$ ;
\ENDWHILE
\STATE $N \gets$ last node in $Q$
\RETURN $N$
\end{algorithmic}
\caption{Generating the Huffman Tree of Characters}
\label{alg:Huffman}
\end{algorithm}

Each element in $S_p$ is assigned a leaf node $N_{c^i_j}$ and then pushed into a priority queue $Q$. The priority is determined by the probability $p_{c^i_j}$, that the lower probability $p_{c^i_j}$ has higher priority in the queue $Q$. Then the algorithm generates Huffman tree by removing the 2 highest priority nodes $N_a$ and $N_b$ from $Q$ and then merge them into a new node $N_c$, which takes the higher priority node of the two removed nodes as the left children and the other as the right children. The priority $p_c$ is determined by the sum of $p_a$ and $p_b$. We repeat this process until there is only one node $N$ in $Q$, then $N$ is the root node of the Huffman tree.

\begin{algorithm}
\begin{algorithmic}
\REQUIRE $N$, $c \gets None$
\STATE $N_l$, $N_r \gets$ children of $N$
\IF{$N_l$ is not leaf node}
\STATE $\{{code}_{c^i_j}\} \gets code(N_l, c+0)$
\ELSE
\STATE $code_{N_l} = c+0$
\ENDIF
\IF{$N_r$ is not leaf node}
\STATE $\{{code}_{c^i_j}\} \gets code(N_r, c + 1)$
\ELSE
\STATE $code_{N_r} = c+1$
\ENDIF
\RETURN $\{{code}_{c^i_j}\}$
\end{algorithmic}
\caption{function $code(N, c)$, Huffman Coding of Characters}
\label{alg:coding}
\end{algorithm}

Given the Huffman tree root node $N$, as shown in Algorithm \ref{alg:coding}, recursively we generate the Huffman code ${code}_{c^i_j}$ for each leaf node $N_{c^i_j}$. The root node $N$ starts with an empty representation $c$, for each node the code of its left children $N_l$ is $c + 0$, and the code of its right children $N_r$ is $c + 1$. We can recursively traverse the tree and give every leaf node a corresponding code ${code}_{c^i_j}$.

\begin{algorithm}[h]
\begin{algorithmic}
\REQUIRE $N$, $vocab\_size$, $depth$, $inner\_size$, $hidden\_size$
\STATE /*Preprocessing before training*/
\FOR{i from 0 to $inner\_size-1$}
\STATE non-leaf $node_i.index \gets i$ 
\ENDFOR
\STATE $Index, Sign, Bias \gets$ zero matrices of $(vocab\_size$, $depth)$
\FOR{each $leaf_i$ $\in$ $N$}
\STATE $path_i \gets$ the path from root $N$ to $leaf_i$
\FOR{each $node_j \in path_i$}
\IF{$node_{j+1}$ is $node_j$'s left child}
\STATE $Index_{ij} \gets node_j.index$
\STATE $Sign_{ij} \gets 1$
\ELSIF{$node_{j+1}$ is $node_j$'s right child}
\STATE $Index_{ij} \gets node_j.index$
\STATE $Sign_{ij} \gets -1$ 
\STATE $Bias_{ij} \gets 1$
\ENDIF
\ENDFOR
\ENDFOR
\STATE /*Forward pass*/
\STATE $ h \gets sigmoid(hidden\_vecs * embedding) $
\STATE $H \gets $ stack $h$ vertically $vocab\_size$ times
\STATE $H \gets H$ is index selected by $Index$ in the last dimension
\STATE $ log\_probs \gets \sum\limits_{column}{\log (Sign * H + Bias)} $
\RETURN $log\_probs$
\end{algorithmic}
\caption{Arbitrary Binary Tree Based H-Softmax}
\label{alg:hsoftmax}
\end{algorithm}

\section{H-Softmax} 
\label{ssec:hs_impl}
In order to vectorize calculations on a binary tree (Fig~\ref{fig:tree}), we need to vectorize paths from root to leaves first. 
For a path, every time it turns left or right, we multiply the probability of this path by $1-\sigma(h)$ or $ \sigma(h) $ in which $h$ is inner product of current node's hidden vector and embedding vector fed into H-Softmax. When the path goes to its leaf node, the accumulated probability of a token is obtained.
Thus, a complete path can be composed of three kinds of path elements: $ 1-\sigma(h)$, $\sigma(h)$, or $1$ which are corresponding to left node, right node, or padding node (dot line in Fig~\ref{fig:tree}). We need the padding node to pad each path to depth of the tree. Finally, the accumulated probability is a product of all of the path elements.

Now, let's focus on these path elements. Remember that $\sigma(h)$ is a variable for each different sample, while path elements' sign (-1, 1, 0) and bias (1, 0, 1) is fixed because the tree structure is fixed. So, we can extract signs and bias of a path into two row vectors $ path\_sign\_encoding_i$ and $ path\_bias\_encoding_i $, then stack them vertically to matrices of $ Sign $ and $ Bias $. Meanwhile, we can stack $h$ from all non-leaf node to one column vector.
Because of $\log$ operation, padding nodes which is expressed by sign 0 and bias 1 is automatically eliminated.

We present above algorithms in Algorithm \ref{alg:hsoftmax}. And we need to emphasize that in practice the intermediate matrix $H$ is index selected so that its column number is reduced from $inner\_size$ to $depth$. Since $depth$ is roughly $\log(vocab\_size)$, this step is crucial to maintain the algorithm in time complexity of $O(vocab\_size * hidden\_size + vocab\_size * \log(vocab\_size))$ instead of $O(vocab\_size * hidden\_size + vocab\_size^2)$ while it's $O(vocab\_size * hidden\_size)$ for vanilla softmax.

\end{document}